\documentclass{aa}
\usepackage[utf8]{inputenc}
\usepackage{amsmath}
\usepackage{amsfonts}
\usepackage{amssymb}
\usepackage{color,soul}
\usepackage{longtable}
\usepackage{graphicx, bm}
\usepackage{booktabs}
\usepackage{tablefootnote}
\usepackage{placeins}
\begin{document}

\title{Activity distribution of comet 67P/Churyumov-Gerasimenko from combined measurements of non-gravitational forces and torques}
\titlerunning{Activity distribution of comet 67P}

\author{N.~Attree \inst{1} \and L.~Jorda \inst{2} \and O.~Groussin \inst{2} \and J.~Agarwal \inst{1} \and R.~Lasagni Manghi \inst{3} \and P.~Tortora \inst{3,4} \and M.~Zannoni \inst{3,4} \and R.~Marschall \inst{5}}

\institute{Institut f{\"u}r Geophysik und extraterrestrische Physik, Technische Universit{\"a}t Braunschweig, Mendelssohnstr. 3, 38106 Braunschweig, Germany (\email{n.attree@tu-braunschweig.de})
\and
Aix Marseille Univ, CNRS, CNES, Laboratoire d'Astrophysique de Marseille, Marseille, France
\and
Alma Mater Studiorum - Universit{\`a} di Bologna, Dipartimento di Ingegneria Industriale, Via Fontanelle 40, I-47121 Forl{\`i}, Italy
\and
Alma Mater Studiorum - Universit{\`a} di Bologna, Centro Interdipartimentale di Ricerca Industriale Aerospaziale, via Baldassarre Carnaccini 12, I-47121, Forl{\`i}, Italy
\and
CNRS, Laboratoire J.-L. Lagrange, Observatoire de la Côte d’Azur, Boulevard de l’Observatoire,
CS 34229 - F 06304 NICE Cedex 4, France
}

\abstract{}
   {Understanding the activity is vital for deciphering the structure, formation, and evolution of comets. We investigate models of cometary activity by comparing them to the dynamics of 67P/Churyumov-Gerasimenko.} 
   {We matched simple thermal models of water activity to the combined Rosetta datasets by fitting to the total outgassing rate and four components of the outgassing induced non-gravitational force and torque, with a final manual adjustment of the model parameters to additionally match the other two torque components. We parametrised the thermal model in terms of a distribution of relative activity over the surface of the comet, and attempted to link this to different terrain types. We also tested a more advanced thermal model based on a pebble structure.}
   {We confirm a hemispherical dichotomy and non-linear water outgassing response to insolation. The southern hemisphere of the comet and consolidated terrain show enhanced activity relative to the northern hemisphere and dust-covered, unconsolidated terrain types, especially at perihelion. We further find that the non-gravitational torque is especially sensitive to the activity distribution, and to fit the pole-axis orientation in particular, activity must be concentrated (in excess of the already high activity in the southern hemisphere and consolidated terrain) around the south pole and on the body and neck of the comet over its head. This is the case for both the simple thermal model and the pebble-based model. Overall, our results show that water activity cannot be matched by a simple model of sublimating surface ice driven by the insolation alone, regardless of the surface distribution, and that both local spatial and temporal variations are needed to fit the data.}
   {Fully reconciling the Rosetta outgassing, torque, and acceleration data requires a thermal model that includes both diurnal and seasonal effects and also structure with depth (dust layers or ice within pebbles). This shows that cometary activity is complex. Nonetheless, non-gravitational dynamics provides a useful tool for distinguishing between different thermophysical models and aids our understanding.}
   

\maketitle

\section{Introduction}

Comets are amongst the most primordial Solar System objects. They formed directly from the protoplanetary disc and survived mostly unaltered for much of their lifetimes in the outer Solar System. They are therefore vital targets for our understanding of planet formation and the history of the early Solar System. Upon entering the inner Solar System, comets are heated by the Sun and undergo activity; that is, ices are sublimated and gas and dust are ejected. Cometary activity poses open questions related to the structure, composition, and thermophysical properties of the nucleus material. This is directly connected to their formation in the early Solar System. Whether cometary nuclei, and by extension planets, formed from the gravitational collapse of clouds of ~centimetre-sized pebbles (as  proposed in \citealp{Blum2017}) or by continual collisional growth \citep{Davidsson2016} has direct implications for the structure and strength of the near-surface material that controls outgassing.

In addition to being directly observable, the outgassing produces a reaction force on the nucleus that can alter its trajectory (as first recognised by \citealp{Whipple50} and described by \citealp{Marsden1973}) and rotation state (see \citealp{Samarasinha2004}). Measuring the changing orbits and spins of comets therefore provides a useful insight into the the micro-physics of the activity mechanism.

Many thermophysical models have been proposed to explain the activity (see recent examples by \citealp{Fulle2019}, \citealp{Gunlach2020}, and \citealp{Davidsson2021MNRAS}), and these can be compared to the outgassing rates of observed comets. In particular, comet 67P/Churyumov-Gerasimenko (67P hereafter) provides an excellent dataset because it was visited by the Rosetta spacecraft between 2014 and 2016.  The spacecraft collected detailed measurements of the size, shape, surface properties, and time-varying rotation state and outgassing of the nucleus. Finding the distribution of activity across the nucleus of 67P that fits the various measurements of the total outgassing rate best (\citealp{Hansen, Marshall2017, Combi2020, Laeuter2020}, etc.) has produced several so-called activity maps (e.g. \citealp{Marschall16, Marschall17, Laeuter2020}, ), which are often expressed as an effective active fraction (EAF) relative to a pure water-ice surface. When examining only the summed total outgassing, however, there is always a degeneracy in the retrieved activity distribution \citep{Marschall2020}, whilst, at the same time, the effects of seasonal changes in insolation and dust cover across the surface of 67P are complicated \citep{Keller2017, Cambianica2021}. Comparing the effects of a model outputted non-gravitational acceleration (NGA) and torque (NGT) to the dynamics of 67P can provide a further constraint on the model parameters and on our understanding of the activity \citep{Attree2019, Kramer2019, kramer2019b, Mottola2020}.

Simple NGA models, such as those by \citet{Marsden1973} and \citet{Yeomans89}, parametrise the acceleration using variables scaled to a general water-production curve, and therefore provide limited insight into the physics of the activity on an individual comet. More complex models (following from \citealp{Sekanina93}) relate the observed NGA and NGT to the outgassing via a thermal model and some distribution of ices or active areas across the nucleus surface. If independent measurements of this distribution and/or the outgassing rate can be made, then cometary masses and spin axes can be measured from ground-based observations, as was achieved for 67P \citep{Davidsson05, Gutierrez05}. Rosetta then provided both the detailed outgassing data mentioned above, as well as precise measurements of the nucleus position and rotation via radio-tracking and optical navigation. As summarised in \citet{Mottola2020}, various attempts have been made to compare thermal models to the NGA and NGT forces of 67P \citep{Keller, Davidsson2022} and to fit its non-gravitational trajectory \citep{kramer2019b}, rotation state \citep{Kramer2019}, and both in combination with outgassing \citep{Attree2019}.

In \citet{Attree2019}, our previous paper on this topic, we used the EAF formalism to fit surface distributions to the observed Earth-comet range (the most accurate component of the comet ephemeris, based on the spacecraft radio tracking), total gas production (measured by ROSINA, the Rosetta Spectrometer for Ion and Neutral Analysis; \citealp{Hansen}), and the change in spin rate ($z$ component of the torque, measured as part of the nucleus shape reconstruction; \citealp{Jorda2016}). We found that a large EAF in the southern hemisphere of the comet, as well as an increase in EAF around perihelion, were needed to fit both the total production measurements and the NGA. However, our model was limited by not considering the other components of the NGT (i.e. the change in the spin axis orientation, as well as its magnitude), and by a rather nonphysical way of splitting the surface into areas of differing activity. Additionally, discontinuities in the cometary heliocentric trajectory reconstructed by the European Space Operations Centre that arose because the NGA was excluded from the operational dynamical model, have complicated the analysis by making it difficult to extract smooth acceleration curves.

\citet{kramer2019b} addressed this problem by performing their own N-body integrations with a model following \citet{Yeomans89} and varying initial conditions. They then fitted a smoothed, interpolated curve to the residuals to extract time-varying NGA curves, but they did not compare them to a full thermal model. In a separate paper \citep{Kramer2019}, the authors did compare a physical thermal model, again using the EAF formalism, to both the rotation rate and axis orientation data. Similarly to our results, their results also required a relatively higher EAF in the southern than in the northern hemisphere, as well as an enhanced outgassing response to insolation around perihelion to fit the data. \citet{kramer2019b} noted that the NGT is much more dependent on the spatial distribution of activity than the NGA.
 
Since then, two additional reconstructions of the Rosetta/67P trajectory have been performed \citep{Farnocchia, 2021EGU}. \citet{Farnocchia} used a rotating-jet model following \citet{Sekanina93} to fit ground-based astrometric observations and radio-ranging measurements before and after perihelion (where the spacecraft NGAs are smaller and the range accuracy is higher). \citet{2021EGU}, on the other hand, used the full Rosetta two-way range and differential one-way range ($\Delta$DOR) dataset, also including low-accuracy data close to perihelion. They tested various NGA models, including a rotating-jet model, and found a best-fit trajectory using an empirical, stochastic acceleration model. Both of these works produced acceleration curves to which a thermal model can be compared.

\citet{Davidsson2022} did just that by comparing the output of a more complex thermal model (NIMBUS; \citealp{Davidsson2021MNRAS}) to the acceleration curves of \citet{Farnocchia} and \citet{kramer2019b}. They found relatively good agreement without fitting, but had to vary several model parameters (e.g. the sublimation-front depth and the gas diffusivity) between the northern and southern hemispheres and pre- and post-perihelion, in order to match the outgassing data. This reinforces the ideas of a hemispheric dichotomy and time-dependent thermophysical properties, and it also demonstrates the complicated nature of trying to model the full thermophysical system of sublimation, gas flow, and dust.

These studies show the usefulness of considering the non-gravitational dynamics. No study has analysed the full six components of NGA and NGT simultaneously, however (we analyse all six here, but only four are included in the formal fitting procedure), and several other weaknesses exist, such as nonphysical surface distributions or complicated descriptions leading to unfitted models. It is pertinent, therefore, to re-examine the full non-gravitational dynamics of 67P with a simple thermal model that can be parametrised in terms of real surface features while being easily compared with more complicated models. This is what we attempt to do here, bearing in mind that the aim is not to find the full description of cometary activity, but a model that adequately describes the data and points towards the underlying physics.

The rest of this paper is organised as follows: in Section \ref{method} we describe how we updated the model of \citet{Attree2019} for use here. In Section \ref{results} we describe three different parametrisations of the surface activity distribution and their results in the model fit. These results are discussed, with reference to a run with the more advanced thermal model of \citet{Fulle2020} in Section \ref{discussion}, before we conclude in Section \ref{conclusion}.

\section{Method}
\label{method}

We followed the method of the first paper \citep{Attree2019} by first calculating surface temperatures over a shape model of 67P ({\it SHAP7}; \citealp{Preusker17}) with a simple energy-balance thermal model and then computing the resulting non-gravitational forces and torques and implementing them in an N-body integration. The model was then optimised by scaling the relative activity of various areas of the shape model up and down, minimising the residuals to the observed datasets: the Earth-comet range (i.e.~the scalar projection of the three-dimensional comet position in the Earth-comet direction, $R$, with $N_{R}=1000$ data points) or the directly extracted NGAs from \citet{2021EGU} (with $N_{NGA}=17000$ data points in each of the three components); the total gas production ($N_{Q}=787$, \citealp{Hansen}); and the spin-axis ($z$) aligned component of the torque ($N_{T_{z}}=1000$, \citealp{Jorda2016}). Additionally, we now also computed the change in the orientation of the rotation axis \citep{Kramer2019} and used this as an output to compare different models.

The thermal model computes the surface energy-balance, taking insolation, surface thermal emission, sublimation of water ice, projected shadows, and self-heating into account (see \citealp{Attree2019} for details). Heat conduction into the nucleus is neglected for numerical reasons, but is small because of the low thermal inertia of the comet \citep{Gulkis2015}. Heat conduction would mainly affect night-time temperatures, which are very low and contribute little to the outgassing (but see the discussion in Section \ref{discussion}). Again for numerical reasons, surface temperatures are calculated roughly once every 10 days for a full 12.4 hour rotation, and the derived quantities are interpolated (see details below) to produce smooth curves over the full mission period of $\text{about two}$ years. Surface temperatures are each computed twice, once assuming an effective active fraction $\rm{EAF}=0$ (i.e.~pure grey-body dust surface), and once with $\rm{EAF}=1$ (i.e.~sublimation from a pure water-ice surface), and the temperatures and sublimation rates are saved. In the fitting process, the pure water-ice sublimation rate is then scaled by a variable EAF and is used, along with the sublimation gas velocity calculated from the zero-ice surface temperature, to compute the outgassing force per facet. The momentum coupling parameter was assumed to be $\eta=0.7$ \citep{Attree2019}. Torque per facet was also calculated here using the ``torque efficiency'' formalism used before \citep{Keller}, where $\bm{\tau}$ is the facet torque efficiency or moment arm, which is a geometric factor that was computed once at the beginning of the run. The use of the higher zero-ice temperature for the gas thermal velocity assumes that the gas equilibrates with the dusty surface, and this means that our derived EAF values may be lower estimates compared with some other thermal models.

The N-body integration was performed using the open-source $REBOUND$ code\footnote{\url{http://rebound.readthedocs.io/en/latest/}} \citep{ReinLiu}, complete with full general relativistic corrections \citep{Newhall} as implemented by the $REBOUNDx$ extension package\footnote{\url{http://reboundx.readthedocs.io/en/latest/index.html}}, and including all the major planets as well as Pluto, Ceres, Pallas, and Vesta. Objects were initialised with their positions and velocities in the J2000 ecliptic coordinate system according to the DE438 Solar System ephemerides \citep{Standish}, with 67P given its initial state vector from the new Rosetta trajectory reconstruction of \citet{2021EGU} (Table \ref{table2}). The system was then integrated forward in time from $t=-350$ to $+350$ days relative to perihelion, using the IAS15 integrator \citep{ReinSpiegel} and the standard equations of motion, with the addition of a custom acceleration, $\bm{a}_{NG}$, for 67P, provided by our model. The Earth-comet range, which is the most accurate component of the comet trajectory, was computed for comparison with the reconstructed trajectory (extracted using the $SpiceyPy$ Python package; \citealp{SpiceyPy}).

A bounded least-squares fit to the residuals was then performed using standard methods implemented in Scientific Python whilst varying the EAF parameters. When forming the overall objective function to be minimised (see Eqns. 9 and 10. in \citealp{Attree2019}), the datasets were weighted by a factor $\lambda$ so that each contributed roughly the same to the overall fit (see Table \ref{table}). The datasets used in all fits were the model outputted total outgassing rate and the $z$ component of the torque, both with $\lambda_{Q} = \lambda_{T_{z}} = 1$. Furthermore, in some fits, we then used the computed Earth-comet range (with $\lambda_{R} = 0.02$), while in others, we directly compared to the three components of the NGA extracted by \citet{2021EGU} in the cometocentric radial-transverse-normal frame (radial to the Sun, $\hat{r}$, tangential to the orbit, $\hat{t}$, and normal to it). In this case, the integration was only performed once at the end to check the Earth-comet range, but the weighting was zero in the fit ($\lambda_{R}=0$), while $\lambda_{\rm{NGA}}=1$. Performing the N-body integration only once speeds the process up by several times, with individual runs taking a few minutes and fits taking up to a day, depending on the parameters. All parameters were interpolated to the observational data sampling-times using the Fourier method described below.

We first confirm that the \citet{2021EGU} accelerations match the real comet trajectory well when they are input into our N-body integration, and they recover the Earth-comet range to within a few hundred metres. This residual, which is most likely the result of the different integration techniques and perturbing bodies we used, is well below the uncertainty of our thermal model runs.

Previously, the $x$ and $y$ components of the torque vector were discarded, but they were now used when we calculated the changes in pole orientation. In principle, the rates of change of the angular velocity ($\bf{\Omega}$) of the comet around its three principal axes can be related (see e.g. \citealp{Julian1990}) to the torque components by
\begin{equation}
\begin{aligned}
I_{x} \dot{\Omega_{x}} = (I_{y} - I_{z})\Omega_{y}\Omega_{z} + T_{x},\\
I_{y} \dot{\Omega_{y}} = (I_{z} - I_{x})\Omega_{x}\Omega_{z} + T_{y},\\
I_{z} \dot{\Omega_{z}} = (I_{x} - I_{y})\Omega_{x}\Omega_{y} + T_{z},
\label{eqn_rot}
\end{aligned}
\end{equation}
where $I_{x}=9.559 \times 10^{18}$, $I_{y}=1.763 \times 10^{19}$, and $I_{z}=1.899 \times 10^{19}$ kg m$^{2}$ are the moments of inertia derived from the shape model assuming a constant density of $538\ \mathrm{kg}\ \mathrm{m}^{-3}$ \citep{Preusker17}, and to the pole orientation right ascension, RA, and declination, Dec, by
\begin{equation}
\begin{aligned}
\dot{\psi} &= \frac{-\Omega_{y} \cos(\psi) - \Omega_{x} \sin(\psi)}{\tan(\theta)} + \Omega_{z},\\
\dot{\phi} &= \frac{\Omega_{y} \cos(\psi) + \Omega_{x} \sin(\psi)}{\sin(\theta)},\\
\dot{\theta} &= \Omega_{x} \cos(\psi) - \Omega_{y} \sin(\psi),
\label{eqn_euler}
\end{aligned}
\end{equation}
via the Euler angles $\phi=\pi/2 + \rm{RA}, \theta=\pi/2 - \rm{Dec}$, and $\psi$. 

In practice, the fact that our model runs over individual rotations separated by gaps means that the torque curves are discontinuous and cannot be directly integrated. We therefore followed the technique of \citet{Kramer2019} and applied a Fourier analysis to the torque curves. The method proceeds by i) extracting the torque over a single rotation as a function of the sub-solar longitude, using \citet{Kramer2019} Eqns. 26, 27, ii) computing the Fourier transform as a function of sub-solar longitude using Eqn. 23, iii) interpolating the Fourier terms as smooth curves over the full Rosetta period; Eqn. 24, and iv) reconstructing the torque at a chosen time by the inverse Fourier transform; Eqn. 25. This allows the calculation of a smoothly interpolated torque value at any given time, $T_{x,y,z}(t)$, for use in the rotation equations (\ref{eqn_rot}).

The set of simultaneous differential equations given by Eqns.~\ref{eqn_rot} and \ref{eqn_euler} was then integrated using standard functions in Scientific Python and the initial conditions $\rm{RA}=69.427^{\circ}, \rm{Dec} = 64.0^{\circ}, \text{and } \psi = 330.703^{\circ}$ at $t=-377.22$ days relative to perihelion \citep{Kramer2019} for the period $t=[-377.22:402.48],$ corresponding to the duration of the Rosetta measurements. The resulting $\rm{RA}(t), \rm{Dec}(t)$ values were not used in the fit due to technical limitations, but were directly compared with the observations as a model output.

\section{Results}
\label{results}

\subsection{Model C}

\begin{figure}
\resizebox{\hsize}{!}{\includegraphics{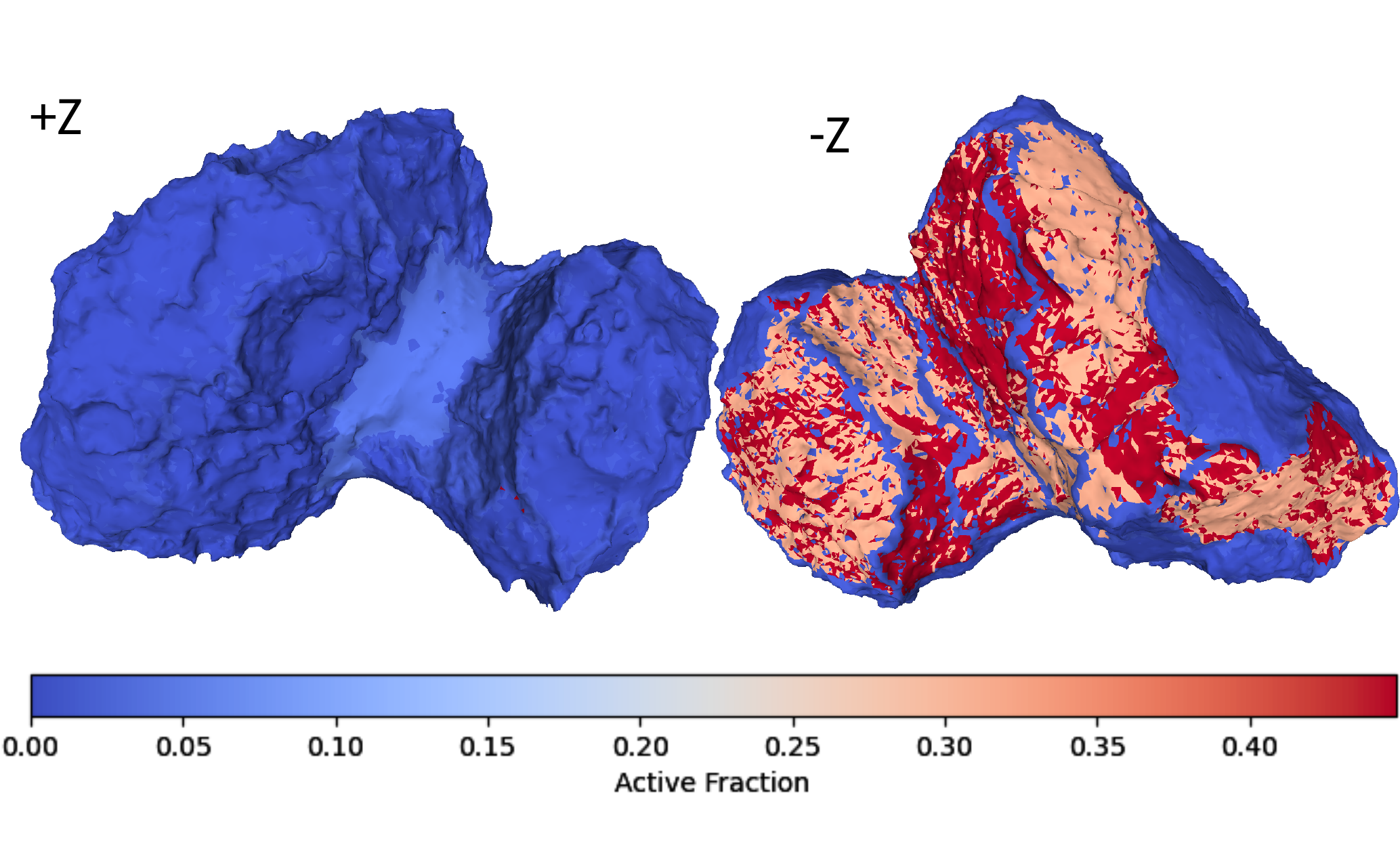}}
\caption{Peak effective active fraction at perihelion for solution C, mapped onto the shape model.}
\label{Map_AcFrac_C}
\end{figure}

We began by rerunning the best-fit model of the previous paper, designated model C in \citet{Attree2019}. This model parametrised the activity distribution by splitting the surface into the 26 regions, defined by \citet{Thomas15} (see their figures for maps), and then grouping them into five super-regions following \citet{Marschall16}(see Figure 4 in \citealp{Attree2019}), before finally splitting the Southern super-region into two (see Figure 17 in \citealp{Attree2019}) and allowing these to vary their EAF with time. With 6 super-regions and the 6 time-variation parameters, there are a total of 12 free parameters in this model. These super-regions consist of region 1, covering the equatorial areas; region 2, covering the base of the comet body and top of the head; the individual regions Hathor and Hapi; and two southern super-regions split on a per-facet basis by the sign of the $z$ component of the torque efficiency (i.e.~south positive with $\tau_{z}>0$ and south negative with $\tau_{z}<0$). This splitting was the only way in which a satisfactory fit to the $z$ torque (i.e.~rotation-rate data) could be achieved, but it remains somewhat artificial. Figure \ref{Map_AcFrac_C} shows the best-fit solution achieved here, mapped onto the shape model. This shows the discontinuous and patchy appearance of the southern super-regions, as well as the north-south EAF dichotomy and activity in Hapi (the light blue area in the northern neck region).

\begin{figure}
\resizebox{\hsize}{!}{\includegraphics{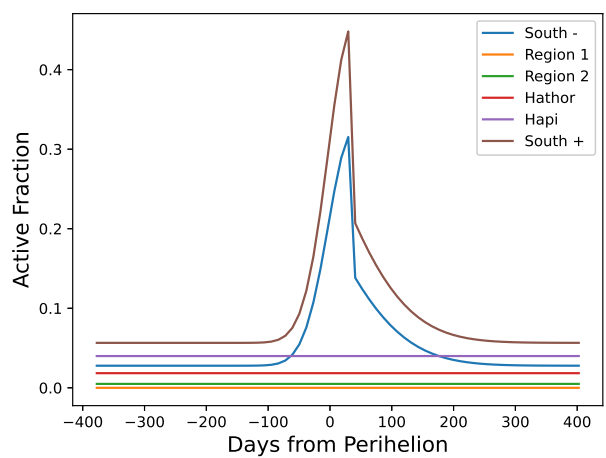}}
\caption{Time-varying effective active Fraction for solution C.}
\label{Plot_AcFrac_C}
\end{figure}

We optimised this model again here and, with a slightly differing procedure for sampling and interpolating the computational output, produced very similar results to before, with no significant improvement in the fit. Next, we instead fit the model directly to the \citet{2021EGU} NGA curves as described above, producing the best-fit solution shown mapped onto the shape-model in Figure \ref{Map_AcFrac_C} (where the values shown are peak EAF, the maximum value for all times), and with time in Figure \ref{Plot_AcFrac_C}. The output is very similar to the previous solution in \citet{Attree2019}, but Figure \ref{Plot_AcFrac_C} shows an even more pronounced spike in EAF around perihelion than before.

\begin{figure}
\resizebox{\hsize}{!}{\includegraphics{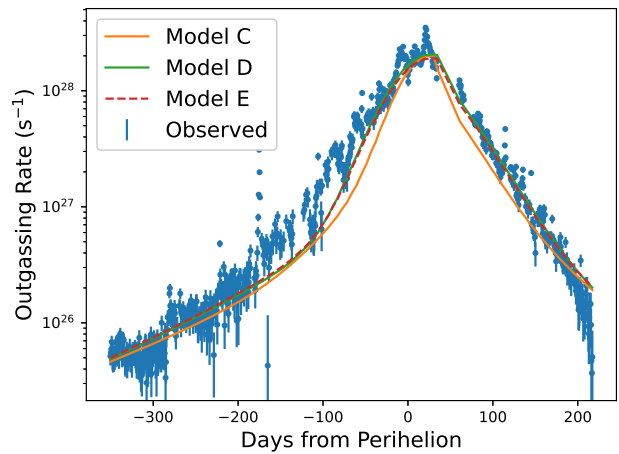}}
\caption{Observed total gas production (ROSINA values from \citealp{Hansen}) compared to solutions C, D, and E.}
\label{Plot_resultsQ}
\end{figure}

\begin{figure}
\resizebox{\hsize}{!}{\includegraphics{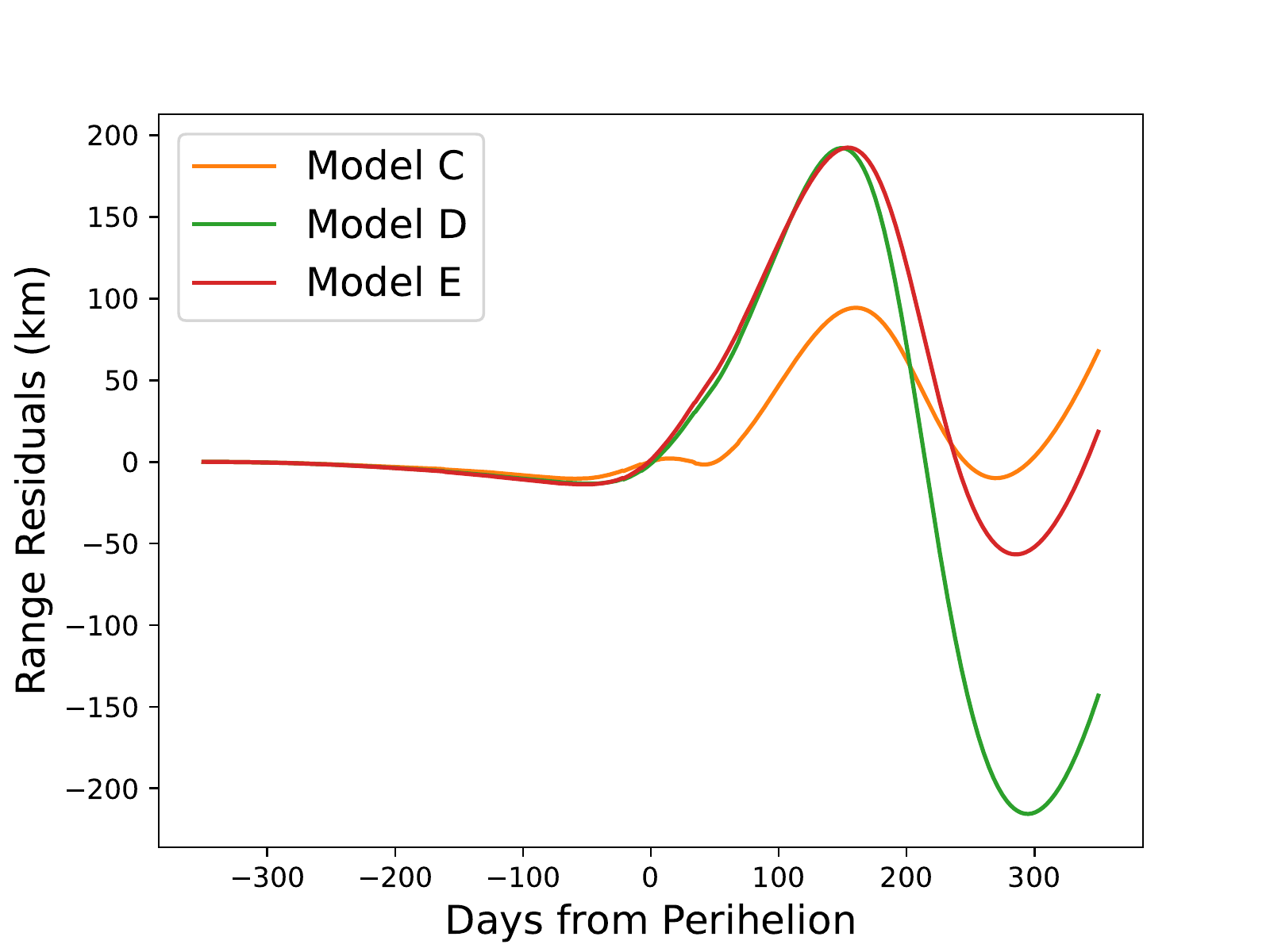}}
\caption{Observed minus computed Earth-comet range, $R$, for solutions C, D, and E.}
\label{Plot_resultsR}
\end{figure}

\begin{figure}
\resizebox{\hsize}{!}{\includegraphics{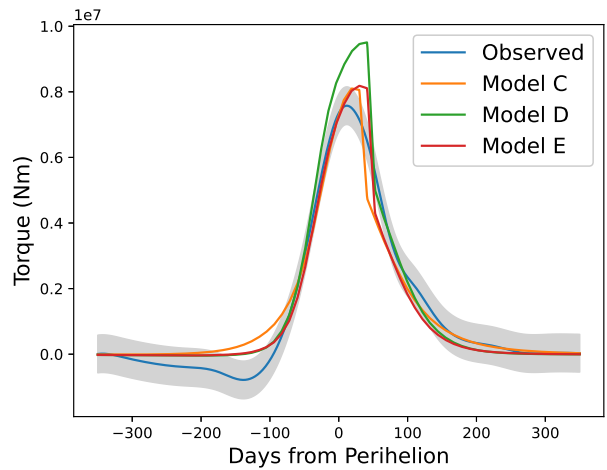}}
\caption{Smoothed observed $z$ component of the torque compared to solutions C, D, and E. The grey area represents the 1$\sigma$ uncertainty (see \citealp{Attree2019} for details).}
\label{Plot_resultsTorque}
\end{figure}

The model fits are shown in the orange curves in Figures \ref{Plot_resultsQ}, \ref{Plot_resultsR}, and \ref{Plot_resultsTorque}, with the fit statistics in the first line of Table \ref{table}. The $z$ torque (Fig. \ref{Plot_resultsTorque}) and total gas production from ROSINA (Fig. \ref{Plot_resultsQ}) are reasonably well fit, with the perihelion peak-values matched, but with a slightly differing shape around the inbound equinox roughly 100 days before perihelion. An improvement in the trajectory fit is attained, with the new RMS residual value of 34 km reduced from the previously achieved 46 km. The shape of the curve is similar.

\begin{figure}
\resizebox{\hsize}{!}{\includegraphics{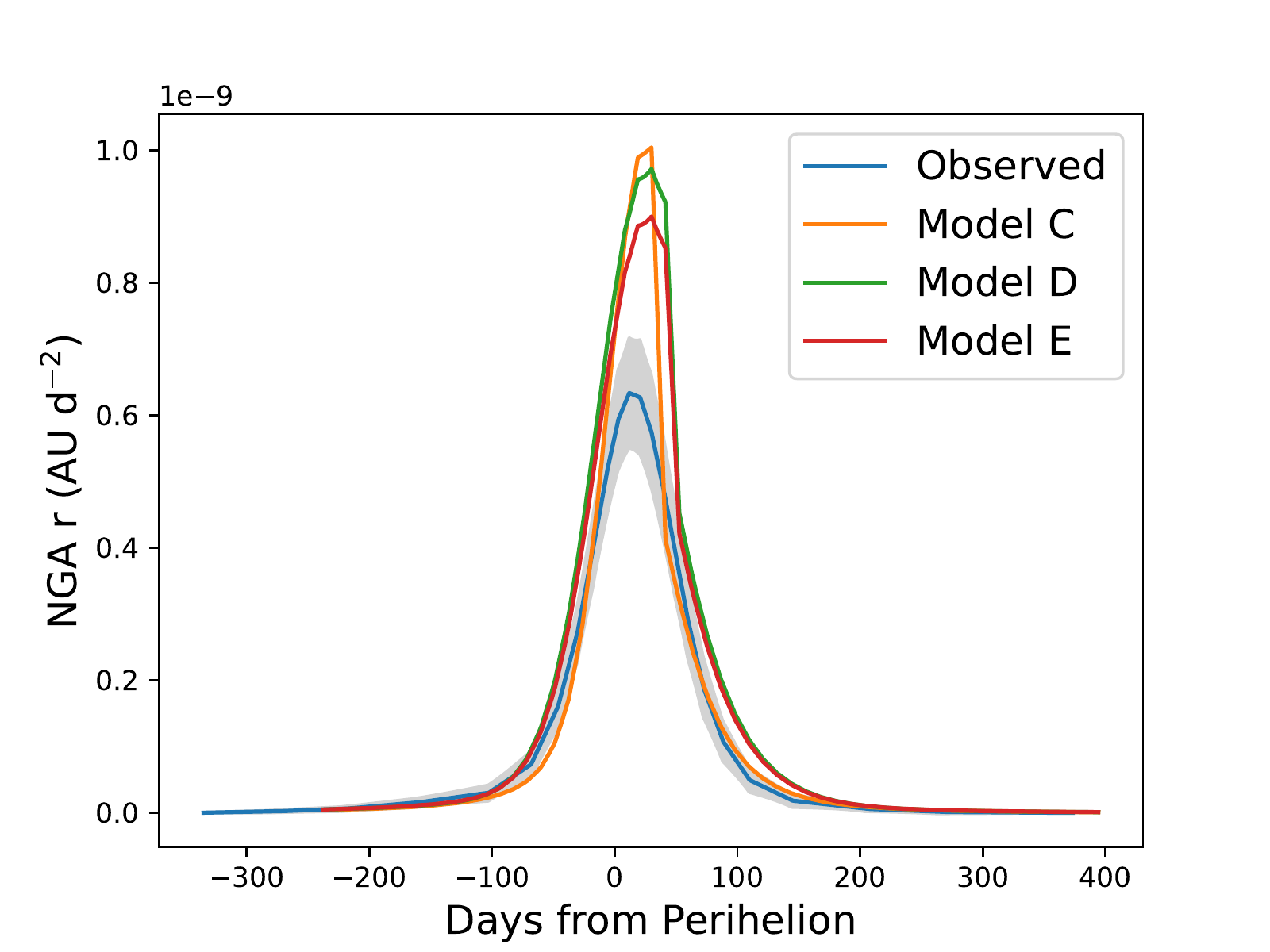}}
\caption{Observed radial acceleration in the comet \textbf{$(\hat{r},\hat{t},\hat{n})$} frame with the 5$\sigma$ uncertainty (from \citealp{2021EGU}), compared to solutions C, D, and E. Higher-order Fourier terms corresponding to daily oscillations are omitted for clarity, but are included in the fit.}
\label{Plot_NGAr2}
\end{figure}

\begin{figure}
\resizebox{\hsize}{!}{\includegraphics{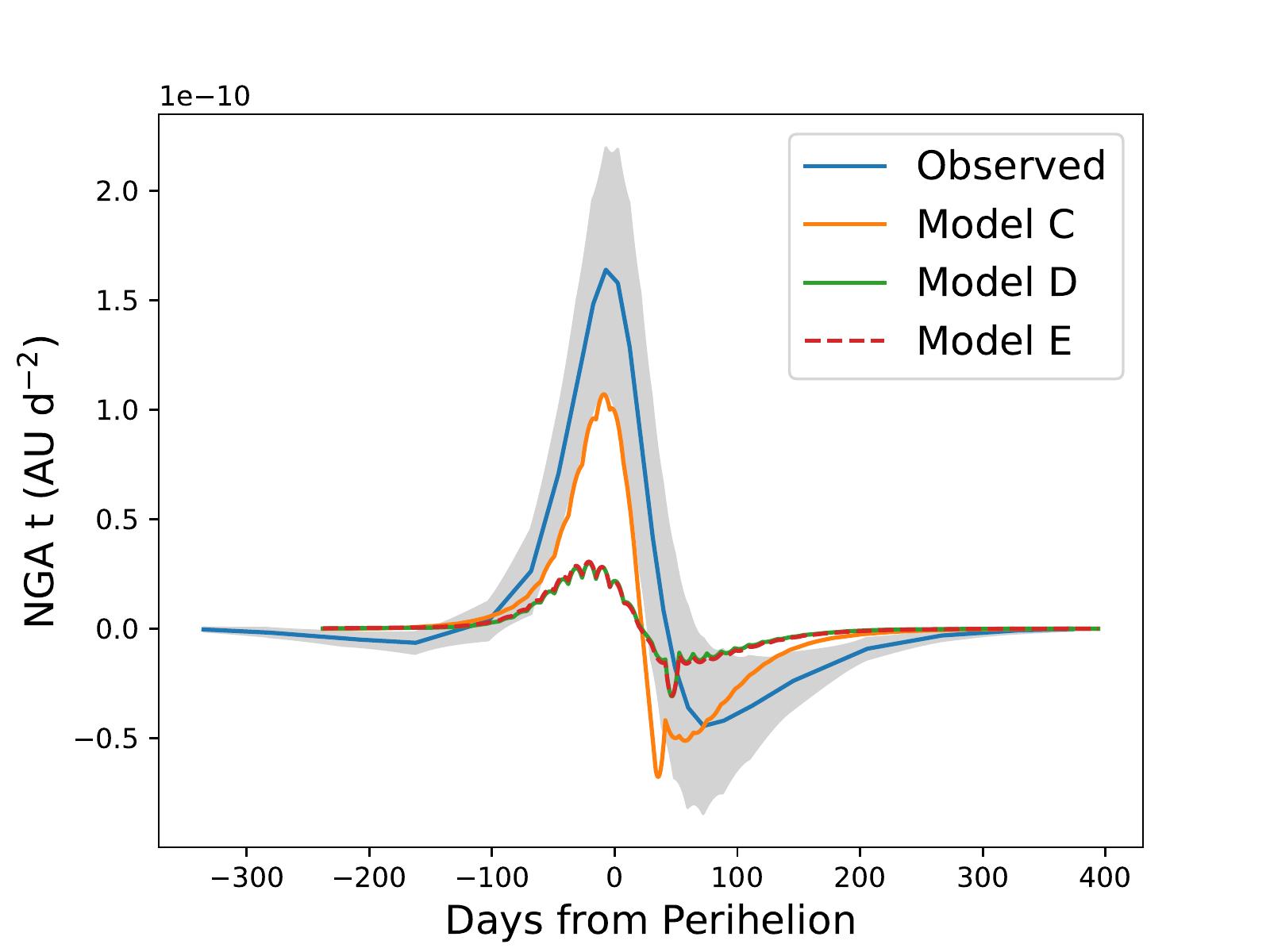}}
\caption{Observed tangential acceleration in the comet \textbf{$(\hat{r},\hat{t},\hat{n})$} frame compared to solutions C, D, and E.}
\label{Plot_NGAt2}
\end{figure}

\begin{figure}
\resizebox{\hsize}{!}{\includegraphics{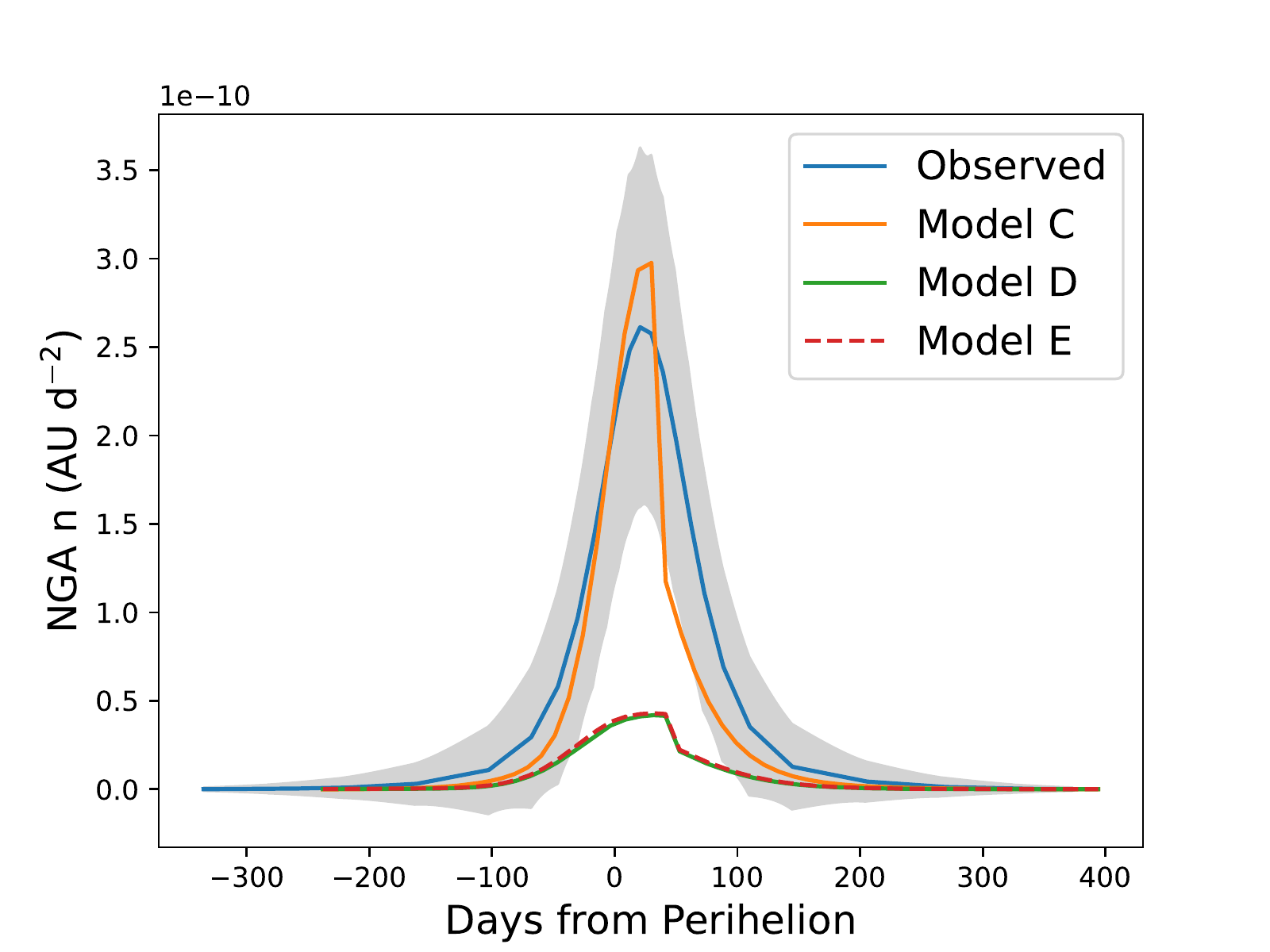}}
\caption{Observed normal acceleration in the comet \textbf{$(\hat{r},\hat{t},\hat{n})$} frame compared to solutions C, D, and E.}
\label{Plot_NGAn2}
\end{figure}

The orange curves in Figures \ref{Plot_NGAr2}, \ref{Plot_NGAt2}, and \ref{Plot_NGAn2} show the individual acceleration curves in the cometocentric $(\hat{r},\hat{t},\hat{n})$ frame compared to the values extracted by \citet{2021EGU}. The radial component makes up the bulk of the acceleration and is reasonably well matched by model C, with the peak value being $\sim50\%$ too high. The normal and tangential components are of smaller magnitude and are reasonably well fit; the secondary, negative peak of the tangential component after perihelion is the worst area of the fit. The remaining 34 km residuals to the observed trajectory most likely stem from our inability to fit this area of the tangential acceleration, combined with the too large radial component peak.

\begin{figure}
\resizebox{\hsize}{!}{\includegraphics{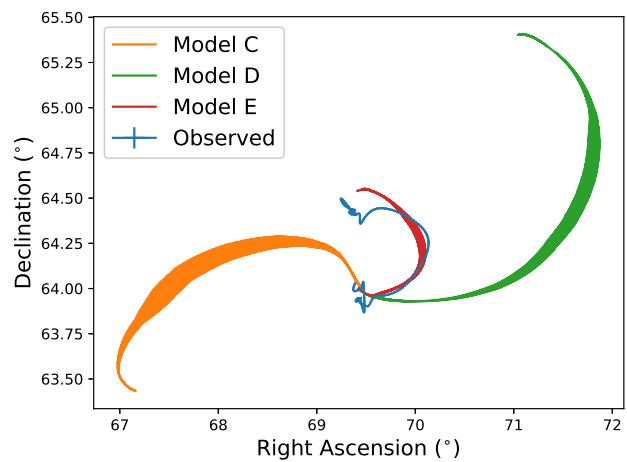}}
\caption{Observed pole orientation (Ra, Dec) compared to solutions C, D, and E. The thickness of the model lines is due to the daily oscillations. Error bars are plotted for the observations, but are small at this scale.}
\label{Plot_resultsRAdec}
\end{figure}

When the pole orientation was calculated, as shown in the orange curve of Figure \ref{Plot_resultsRAdec}, it was a very poor fit to the data, moving off in the opposite direction to the observed changes. This demonstrates that the problem is ill-posed with multiple solutions, and it also highlights the usefulness of including the RA, Dec pole measurement to help distinguish between different models that fit the other data equally well.

\begin{table*}
\caption{Fit statistics for best-fit models C, D, and E, and the two unfitted versions of F.}
\begin{center}
\begin{tabular}{lccccccccccc}
Solution & \multicolumn{4}{c}{Weighting} & \multicolumn{7}{c}{$\chi^{2}$} \\
 & $\lambda_{Q}$ & $\lambda_{T_{z}}$ & $\lambda_{R}$ & $\lambda_{\rm{NGA}}$ & R & Q & T$_{z}$ & NGA$_{r}$ & NGA$_{t}$ & NGA$_{n}$ & Obj \\
\hline
C   & 1 & 1 & 0 & 1 & 34.1 & 4.53 & 1.36 & 1.18 & 1.32 & 0.44 & 1.20 \\ 
D   & 1 & 1 & 0.02 & 0 & 88.8 & 3.60 & 1.10 & 2.00 & 1.60 & 0.90 & 2.35 \\
\bf{E}   & \bf{1} & $\bf{1}$ & \bf{0.02} & 0 & \bf{83.4} & \bf{3.75} & \bf{0.77} & \bf{1.78} & \bf{1.58} & \bf{0.89} & \bf{2.22} \\
F dust SH & - &- &- &- & 324.5 & 4.62 & 2.09 & 4.12 & 1.71 & 1.01 & - \\
F ice SH  & - &- &- &- & 459.2 & 5.64 & 3.02 & 2.22 & 1.63 & 1.00 & - \\ 
\end{tabular}
\tablefoot{Model E is highlighted as the preferred solution. The model outputs (water production rate, $z$ component of NGT, and the three components of NGA) are compared to the observations, producing the $\chi^{2}$ statistics, which are then weighted according to the $\lambda$ values and combined in the objective function (Eqns. 9 and 10. in \citealp{Attree2019}) to produce the combined fit statistic Obj. All values are dimensionless, although the range values $R$ correspond one-to-one to kilometers.}
\label{table}
\end{center}
\end{table*}

\subsection{Model D}

We now proceed with a more physically meaningful model. This was constructed using the list of 71 sub-regions defined in \citet{Thomas2018} (see the reference for maps of their location). We again created super-regions by collecting these sub-regions, but this time, by placing them into one of the five morphological categories of \citet{Thomas15}: `dust-covered terrains' (Dust for short), `brittle materials with pits and circular structures' (Brittle), `large-scale depressions' (Depression), `smooth terrains' (Smooth), and `exposed consolidated surfaces' (Rock). The sub-regions were assigned according to their descriptions in the table in \citet{Thomas2018}. A few ambiguous examples were tested in both the categories to which their descriptions could apply, without altering our results significantly. The Rock and Smooth terrain types both cover significant areas of the southern hemisphere and following the results of the first paper, we therefore allowed their EAFs to vary with time in the same way as for model C. The facets in each super-region all have the same EAF (either constant or time-varying), regardless of the hemisphere in which they are located. With five regions and 6 time-variation parameters, there are 11 parameters in total for this model, designated `model D'. 

\begin{figure}
\resizebox{\hsize}{!}{\includegraphics{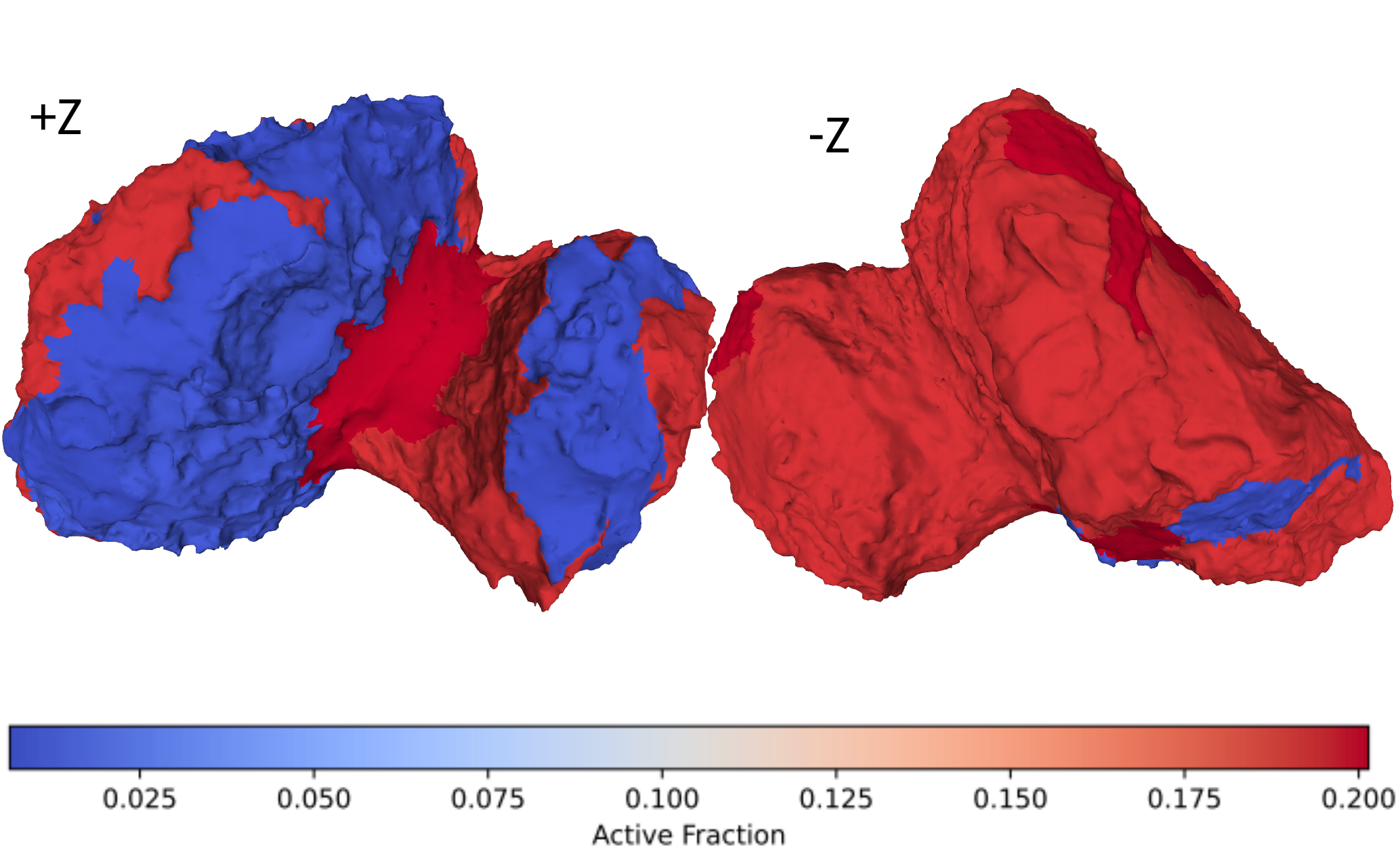}}
\caption{Peak effective active fraction at perihelion for solution D, mapped onto the shape model.}
\label{Map_AcFrac_D}
\end{figure}

\begin{figure}
\resizebox{\hsize}{!}{\includegraphics{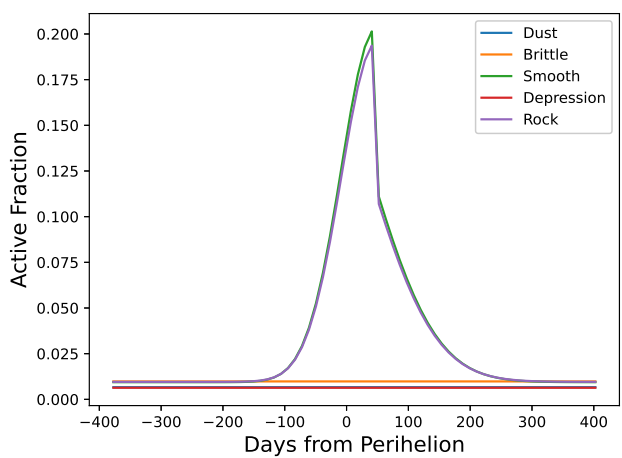}}
\caption{Time-varying effective active fraction for solution D.}
\label{Plot_AcFrac_D}
\end{figure}

Figure \ref{Map_AcFrac_D} shows the peak activity in our best-fit solution for model D mapped onto the shape model, and Fig.~\ref{Plot_AcFrac_D} shows the time variation. High activity is again favoured in the southern hemisphere, with the Rock and Smooth regions seeing much higher activity than the Dusty, Brittle, and Depression regions, especially around perihelion.

Model D is shown as green curves in Figures \ref{Plot_resultsQ} - \ref{Plot_resultsRAdec}. The fit statistics are again shown in Table \ref{table}. This model produces a similar, if slightly improved, fit to the total outgassing measurements, while slightly degrading the trajectory and rotation-rate fits compared to model C. The reasons for the poorer trajectory fit can be seen in the acceleration curves in Figures \ref{Plot_NGAr2}, \ref{Plot_NGAt2}, and \ref{Plot_NGAn2}. The modelled radial component of the acceleration is still slightly too large when compared to the observations, while the tangential and normal components are now much worse than before, with the curves roughly the correct shape, but too small in magnitude. An attempt to fit model D directly to the accelerations did not improve the trajectory, and the individual super-region NGA curves showed no obvious combination that would fit the accelerations better.

 Figure \ref{Plot_resultsRAdec} shows that model D additionally fails to reproduce the observed changes in pole direction. However, the curve now goes in the correct direction, but with a magnitude that is too large compared to the completely incorrect prediction of model C. This suggests that the more physically meaningful model has merit, despite the degraded trajectory fit, and it motivated us to make further adjustments to try and fit all the data below.

\subsection{Model E}

\begin{figure}
\resizebox{\hsize}{!}{\includegraphics{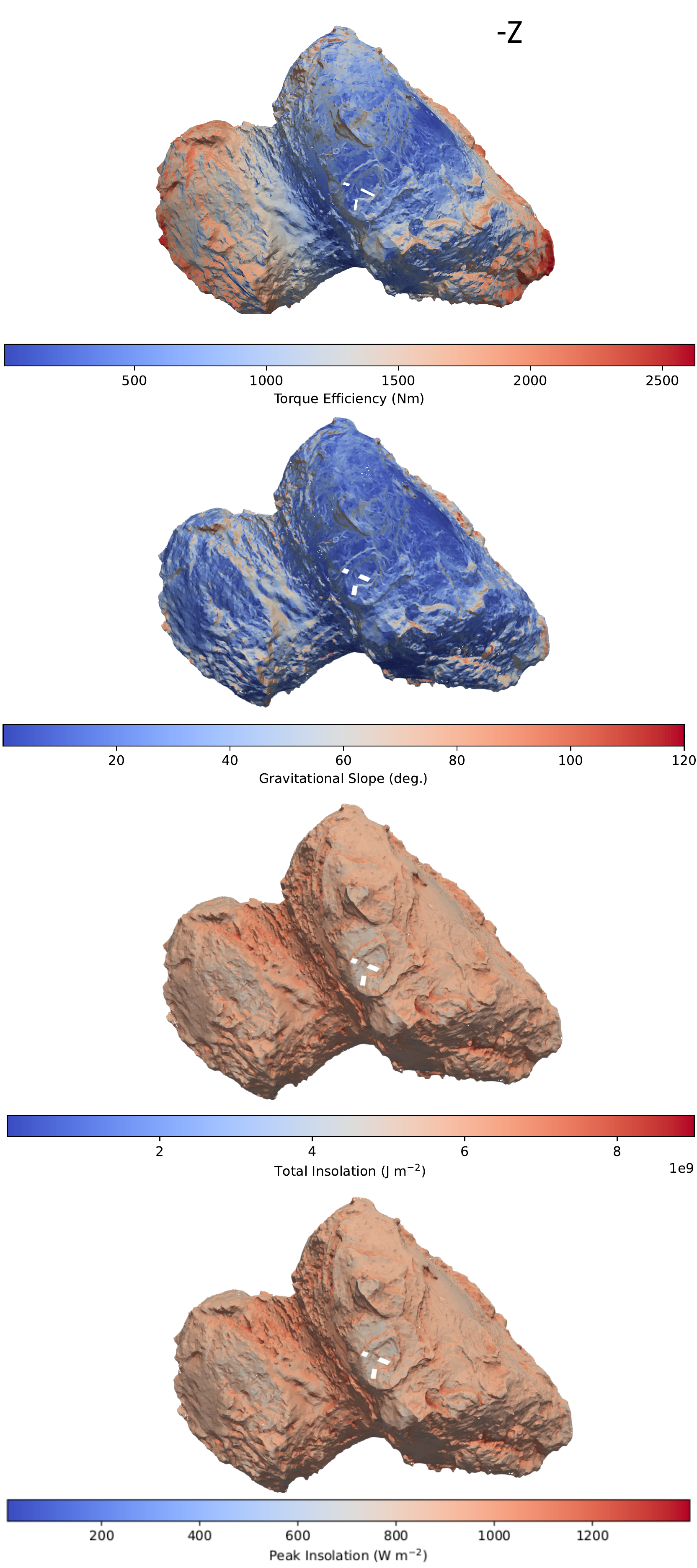}}
\caption{Various datasets mapped onto the southern hemisphere of the comet. From top: Modulus of torque efficiency ($|\tau|$), a geometric factor as described in the text; gravitational slope, i.e. the angle between facet normal and local gravity vector; total integrated insolation; and peak insolation. The three white lines indicate the direction of the $-r$, $-t$, and $-n$ vectors, averaged over one rotation period at perihelion, i.e.~the time-averaged directions towards the Sun, `backwards', and `down' in the orbital frame of the comet.}
\label{Map_torques}
\end{figure}

Because model D fits most of the data well but increasingly fails with the magnitude of the pole direction changes, we sought to modify it by adjusting the NGT. Specifically, in order to fit all the data, the comet must produce a smaller amount of non axial-aligned torque ($x$ and $y$ components), while the rest of the torque and accelerations remain the same. We achieved this in model E with another, somewhat artificial, splitting of the Rock super-region into two super-regions based on their torque contributions. This splitting was performed on a sub-region basis, rather than on the per-facet basis of model C, in order to produce contiguous areas that allowed us to see the general trends in activity across different parts of the comet surface. The modulus of the torque efficiency ($|\tau|$) was first calculated for each facet (top left in figure \ref{Map_torques}) before the area-weighted mean for each sub-region was calculated and the Rock super-region was split into `low torque' ($|\tau|$ lower than the median sub-region value) and `high torque' ($|\tau|$ greater than the median value). Both of these super-regions were allowed to vary with time, leaving a total of 13 free parameters.

\begin{figure}
\resizebox{\hsize}{!}{\includegraphics{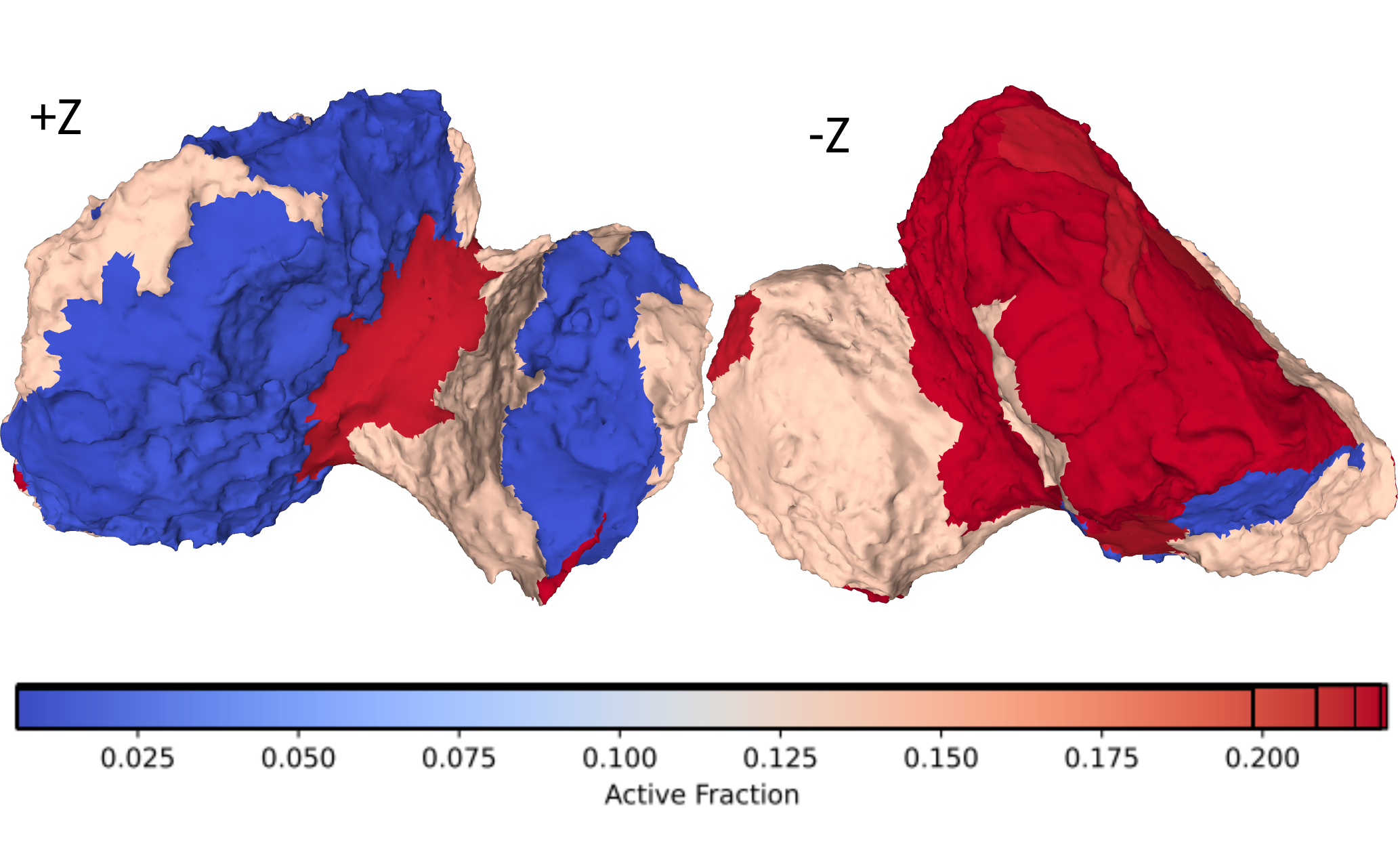}}
\caption{Peak effective active fraction at perihelion for solution E, mapped onto the shape model.}
\label{Map_AcFrac_E}
\end{figure}

\begin{figure}
\resizebox{\hsize}{!}{\includegraphics{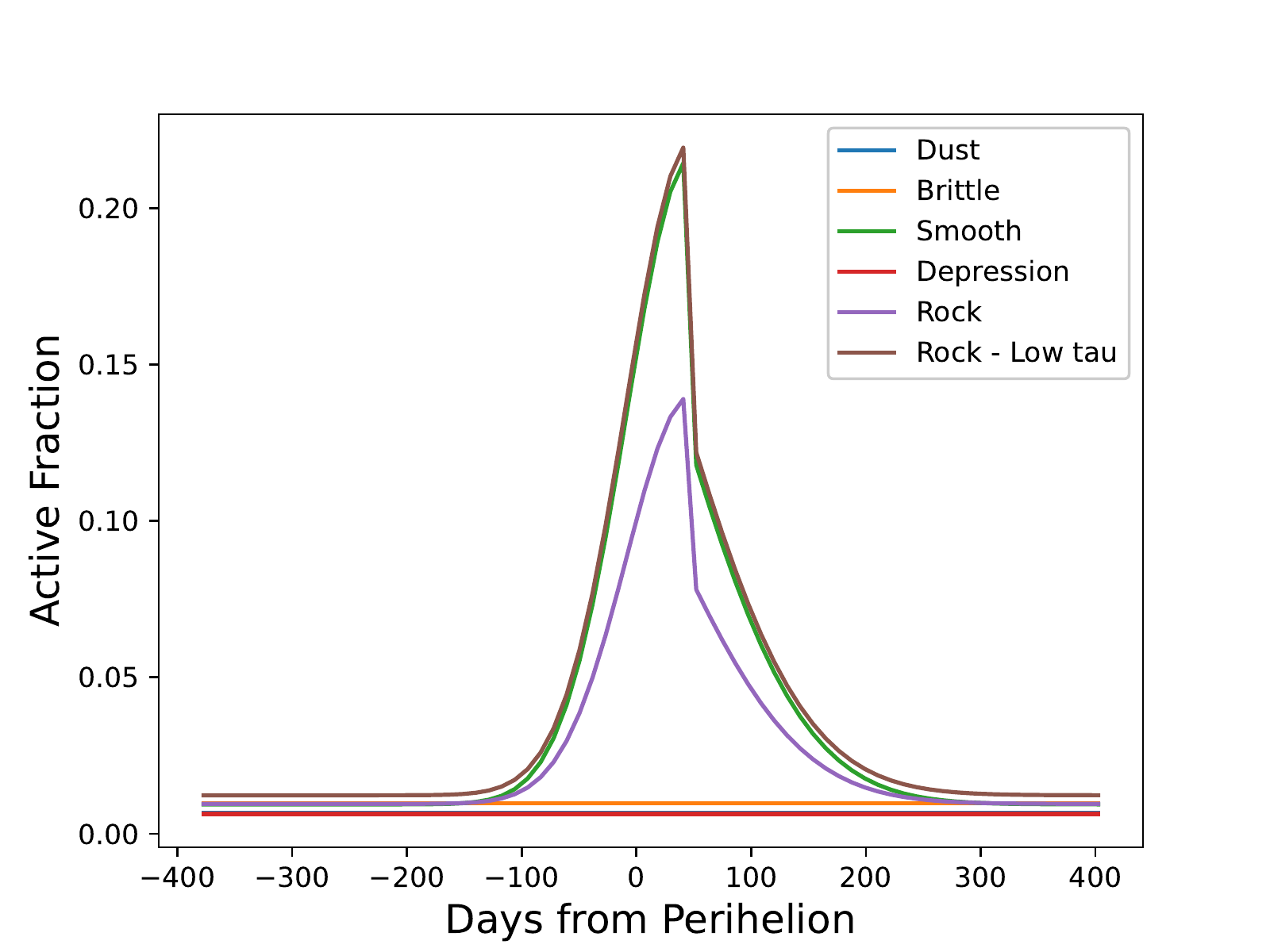}}
\caption{Time-varying effective active fraction for solution E.}
\label{Plot_AcFrac_E}
\end{figure}

Figures \ref{Map_AcFrac_E} and \ref{Plot_AcFrac_E} show the best-fit solution. This was found by manually adjusting the optimised solution by eye to match the pole-direction data. The results are very similar to those of model D, except that the regions of rocky terrain with high torque efficiency are reduced to an intermediate value of activity, between that of the rest of Rock and the other terrain types. The red curves in Figures \ref{Plot_resultsQ} - \ref{Plot_resultsRAdec} show that this adjustment has little effect on the trajectory, production, and rotation-rate fits, but now produces an excellent match to the pole-direction data as well. Thus, model E represents our best-fit solution overall.

When the acceleration curves are considered in detail, model E fails to reproduce the tangential and normal components in the same way as model D. The peak radial acceleration is slightly reduced, however, resulting in a slightly better trajectory fit than for model D. We once again sought improvements in the acceleration by fitting directly to the curves, as well as examining the acceleration produced by individual regions, but no overall better fit was found. Every improvement in the acceleration curves led to a corresponding degradation in the rotation fits.

\section{Discussion}
\label{discussion}

Our best-fit model overall is model E. This model is based on a splitting of the surface according to morphological unit types, with an artificially imposed further splitting according to torque efficiency and a time-varying EAF. A number of trends can be seen across all the solutions, however, which we discuss now, before we return to the interpretation of model E.

In common with the previous results \citep{Attree2019}, all models firstly require a higher EAF in the southern than the northern hemisphere, as well as an EAF that increases around perihelion. This increase in activity, over and above the increase expected with heliocentric distance, is a common result in the literature \citep{Keller, Kramer2019, Davidsson2022} and implies a non-linear outgassing response to insolation. High activity at perihelion is needed to fit the maximum outgassing rate as well as the sharp peak in acceleration, which is mostly contained in the radial component.

Non-gravitational torque, as expressed in the period and spin-axis changes, is much more dependent on the exact spatial distribution of activity (as also found by \citealt{kramer2019b}), especially within this very active southern hemisphere. For example, the correct magnitude of the pole-direction fit is achieved in model E by distributing the activity around the southern hemisphere in a specific way: high activity in regions with low torque efficiency around the south pole, with lower activity in areas with a high torque efficiency, such as towards the extremities of the nucleus and parts of the head. This agrees well with the distribution seen in \citet{Kramer2019} (see their Figs.~9 and 10). As shown in Figure \ref{Map_torques}, these low-torque areas and physical parameters, such as the total amount or peak of insolation received or the gravitational slopes, do not appear to be correlated. The fact that morphologically similar and similarly insolated regions on the head and body show differing levels of activity may imply compositional differences between the two lobes of the nucleus, as suggested by comparisons of region Wosret with the Anhur and Khonsu regions by \citet{Fornasier2021}.

When the seasonal orientation of the comet is considered alongside the acceleration curves, the reasons for the differences between the trajectories of models C, D, and E become clear. The large magnitudes of the normal and tangential acceleration peaks in model C come from the extreme activity ratio of the south polar regions and elsewhere: At perihelion, when the outgassing is at a maximum, the comet orientation is such that the southern hemisphere most often points `downwards' (in the negative direction in the orbital plane, $-\hat{n}$), towards the Sun ($-\hat{r}$), and `backwards' (along the negative of the orbital velocity vector $-\hat{t}$). This is shown in Fig.~\ref{Map_torques} by three vectors, indicating the time-averaged direction of  $\langle-\hat{r}, -\hat{t}, -\hat{n}\rangle$ over one comet rotation at perihelion. As the comet rotates, the unit vectors sweep over its surface, but as a result of the spin-axis orientation at this time, the southern hemisphere points in the indicated direction on average. Thus, the net outgassing force from the southern hemisphere produces a strong positive peak in all three of these components, as seen in the data. Meanwhile, any outgassing from other areas of the comet produces acceleration in different directions, reducing the net positive peaks. This is the case in models D and E (and \citealp{Kramer2019}, etc.), where there is some activity in areas that are not aligned south, meaning that part of the acceleration is in other directions and that the net positive normal and tangential forces are reduced (green and red curves in Figs.~\ref{Plot_NGAt2} and \ref{Plot_NGAn2} compared to orange). The radial peak (Fig.~\ref{Plot_NGAr2}) is less reduced because most outgassing is directed towards the Sun, even in areas that are not aligned south.

When the pole direction is fit, which is dependent on the $x$ and $y$ components of the NGT, however, activity is preferred everywhere, or at least in a less extreme dichotomy than in model C. If the torque distribution in the south-facing regions alone could be adjusted to match the overall, correct, torque distributions of models D and E, then the solutions could be reconciled. However, figures \ref{Map_torques} and \ref{Map_AcFrac_C} show that the correlation between the z component of torque efficiency and its total modulus in the southern hemisphere is complicated, meaning that any adjustment to the pole direction ($x$ and $y$ torque components) will also affect the rotation rate ($z$ component). Any increase or decrease in the perihelion activity of south-facing regions will also strongly affect the acceleration. For this reason, improvement of the acceleration or trajectory fit always degrades the pole direction fit and \textup{}vice versa; the facets controlling NGA and NGT are spatially correlated.

At one instant in time, the non-gravitational torques and accelerations will always be correlated by the spatial pattern described above. However, the total torques and accelerations integrated over some period (e.g. one rotation) may not necessarily be so correlated. For example, torque is evaluated in the body-fixed frame, so that it is independent of the particular orientation of the comet at any one time. The net acceleration vector, on the other hand, depends on this orientation with respect to the Sun and on the heliocentric coordinate frame, and it will vary over a cometary rotation (i.e.~the non time-averaged version of the vectors shown in Fig.~\ref{Map_torques} will rotate around the shape model in the body-fixed frame). In this way, the acceleration per facet integrated over one rotation period will be sensitive to both the total outgassing from the facet over that period and to its temporal variation, whereas the torque will only be dependent on the total outgassing.

A possible way to optimise the fitting to the heliocentric orbit without deteriorating the fit to the rotation-axis orientation and period might then be to redistribute the activity variation with local time. The idea of a lag angle between the peak insolation and peak diurnal activity has indeed been invoked in the past (see e.g. \citealp{Davidsson04}), with recent work suggesting that water activity might peak at 20$^\circ$ \citep{Pinzon2021, Farnocchia} or even 50$^\circ$ \citep{kramer2019b} post-noon, with the latter lag angle varying with time and being undetected before perihelion. Such a lag angle would depend on the thermal inertia and the depth at which water sublimates, making it complicated to model. Additional enhanced activity may also arise at the morning terminator due to sublimation of frost from the night.

CO$_2$ emissions, which have not been considered here, may also have a different local-time distribution. \citet{Pinzon2021} reported a peak at the evening terminator. \citet{Davidsson2022} suggested that CO$_2$ produces little NGA, due to both its small outgassing rate compared to H$_{2}$O and a smoother diurnal variation from a deep sublimation depth and large lag-effect, leading to force in all directions and a cancelling out of the net acceleration. CO$_2$ activity distributed in a specific way, however, might still lead to a net torque, resulting in the required splitting of the torque and acceleration, although it would, admittedly, have to be quite a specific distribution. \cite{Gerig2020} reported that about 10\% of total dust emission originates from the night side, which may well be driven by CO$_2$ emission, while the peak perihelion outgassing rate is roughly one order of magnitude lower than the rate for water \citep{Laeuter2020}.

Clearly, a more realistic thermal model, including thermal inertia as well as possibly the emission of CO$_2$, is needed to fully reconcile the observed outgassing, accelerations and torques. Below, we briefly analyse the results of a recently published thermal model based on \citet{Fulle2020}. This does not include a local time-lag or CO$_2$ emission, but offers an interesting comparison with and extension of the surface energy-balance models discussed above. 

\begin{figure}
\resizebox{\hsize}{!}{\includegraphics{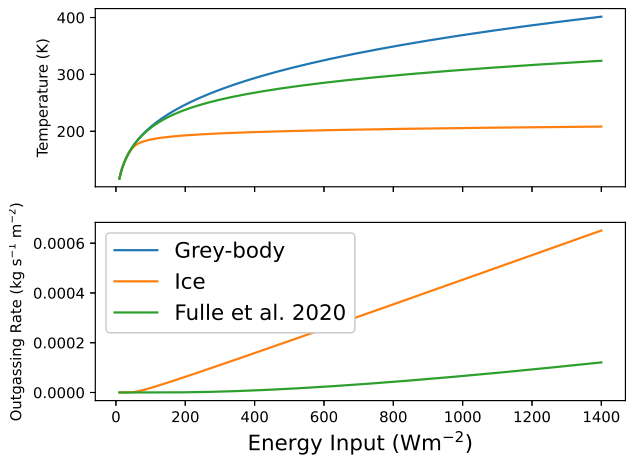}}
\caption{Outputs of the pebble model of \citet{Fulle2020}. Top panel: Surface temperature as a function of energy input for EAF $=0$ grey-body and EAF $=1$ pure-ice surfaces as well as the pebble model. Bottom: Outgassing rate for the pure-ice and the pebble model.}
\label{Plot_TempModels}
\end{figure}

The model of \citet{Fulle2020}, called model F here, assumes a material made of water-containing centimetre-sized pebbles, in which a constant energy balance is maintained between the insolated surface and ice sublimating in the interior of the pebbles. This leads to a set of four differential equations that must be solved simultaneously for each time and facet, instead of the normal surface energy-balance equation. The rest of the code runs as before, with the slight complication that we cannot calculate self-heating in a self-consistent way due to a technical limitation, as it relies on an iteration between facets. We therefore calculated two model F solutions: one solution in which the self-heating per facet was calculated from a pure-ice surface, and another with a pure-dust surface. These two energy inputs bracket the full solution, whose surface temperature (and therefore self-heating term) is intermediate between a pure-ice and a pure-dust grey-body surface (Figure \ref{Plot_TempModels}). The figure also shows that the outgassing rate in the \citet{Fulle2020} model is significantly reduced from that of a pure-ice surface and has a distinctly non-linear shape, ranging between effective active fractions of EAF$ \sim0-20\%$ as a function of insolation.

Figure \ref{Plot_resultsFQ} shows the resulting gas production curve evaluating model F on the shape model, showing that the model of \citet{Fulle2020} can naturally reproduce the high perihelion outgassing rates without the need for an effective active fraction that varies with time. This confirms the results of \citet{Ciarniello2021}.

Figure \ref{Plot_resultsRF} shows the trajectory result obtained with model F, while Figures \ref{Plot_resultsTorqueF} and \ref{Plot_resultsRAdecF} show the torque and pole-direction curves. For a model without any fitting, the results agree reasonably well with the data, although the magnitude of the pole-direction changes are again too large, and the trajectory fit and $z$ torque are not as close as in our best models (see Table \ref{table} for fit statistics).

Figures \ref{Plot_NGArF}, \ref{Plot_NGAtF}, and \ref{Plot_NGAnF} show similar results to before for the accelerations: The overall magnitude of the radial component is approximated well, but the peaks of the tangential and normal accelerations are, again, much too small. The radial acceleration is also not as peaked around perihelion as the observations, while its maximum is closer to perihelion than the observed, delayed peak.

The implications for the pebble-based thermal model are similar to those for the other models. A strong enhancement in activity in the southern hemisphere is needed to fit the narrowly peaked acceleration curves. In model F this is partially provided by the non-linear insolation response, but it is clear that an enhancement beyond even this, or possibly a reduction in activity in other areas, is required. Potentially, this could come from dust fallout from the intensively active southern onto the equatorial and northern regions, quenching them around perihelion.

Finally, experiments in which outgassing in different sub-regions was scaled up and down relative to model F (i.e.~that reintroduced a kind of effective active fraction, but with a different magnitude) also showed a similar response. The large magnitude of the pole-direction change could be reduced by decreasing activity in the high-torque areas, as in solution E, while the trajectory fit could not be improved without degrading the three torque components. This shows that although the pebble model of \citet{Fulle2020} is an improvement over a simple surface energy-balance model, it is still not a complete description of the surface activity distribution of the comet. An even more complex thermal model, possibly requiring time-varying dust fallout as well as thermal inertia and CO$_{2}$, is still required for a fuller description.

\section{Conclusion}
\label{conclusion}

We adjusted a simple thermophysical model to match the combined total outgassing rate and all six components of its resulting non-gravitational forces and torques observed by Rosetta at comet 67P. We parametrised the model in terms of different EAF relative to a pure water-ice surface, and linked their distribution to different terrain types on the comet. We also compared our results to the more complicated thermal model of \citet{Fulle2020}.

Firstly, the results of the fitting confirm the hemispherical dichotomy in relative activity levels (also seen by \citealp{Keller, Kramer2019, Davidsson2022}). The EAF of the southern hemisphere of 67P at perihelion is roughly an order of magnitude larger than that of the northern hemisphere. This increase in relative activity with heliocentric distance (over and above the geometric effect) leads to the steep power-law rise in total outgassing and implies a non-linear response of the surface to insolation. This response arises naturally from the model of \citet{Fulle2020}, which assumes a pebble structure for the nucleus. It might also be caused or enhanced by changes in the thickness of an inert dust-layer resulting from devolatilisation or redistribution of ejected particles (so-called `airfall'), however.

Secondly, for the first time, we correlated differences in responses to insolation with the different terrain types observed on 67P \citep{Thomas15}. We found a good match to most of the Rosetta dataset (total outgassing, NGA, and rotation-rate changes) by doing this. Consolidated Rocky terrains (mainly seen in the southern hemisphere) have the highest relative activity, alongside `smooth' areas in Imhotep, Anubis, and Hapi (\citet{Longobardo19} also report more primordial `fluffy' particles detected by the GIADA instrument over our Rocky consolidated material). Areas with dusty airfall deposits, such as Ma'at and Ash, as well as the floors of the two large depressions (Hatmehit and Aten) and the brittle terrain (mostly located in Seth), have lower activity. These spatial distributions of EAF resemble previous results \citep{Marschall16, kramer2019b}, but are associated with the morphological terrain types for the first time here. Physically, this probably relates to the thickness of the dust covering, with depressions and dusty regions covered in a thick layer of inert fallback material, compared to the relatively volatile-rich exposed consolidated terrain. High activity in the smooth regions such as Hapi (as also noted by \citealp{Marschall16, Fulle2020}) would then represent volatile-rich airfall, which has remained wet during its flight in the coma and stay in the new location, due to local seasonal conditions.

However, this interpretation is complicated by two factors. Firstly, the fact that most consolidated terrain is located in the southern hemisphere, combined with the fact that as a result of the particular seasonal and orbital configuration of 67P, activity here dominates total outgassing, NGA, and NGT. This means that it is difficult to determine the interplay between the intrinsic factors (e.g. the different surface types or compositions) and the extrinsic factors (insolation pattern determined by seasonal effects). The two are indeed likely linked, and the feedback between insolation and dust-cover drives the relative appearance of the two hemispheres.

Secondly, in order to fit the pole-axis orientation data in particular, an additional splitting of activity is needed (NGT is, in general, much more sensitive than NGA to spatial activity patterns). Lower activity is found in some of the extremities of the body, and particularly on the head in the Wosret region, relative to the regions close to the south pole at the boundary of body and neck, even though these regions are not morphologically different or exposed to particularly different patterns of insolation. This is the case both for the basic thermal model and the model of \citet{Fulle2020} that otherwise improves on it. This may imply a compositional or structural difference between the two lobes of the comet (as suggested by \citealp{Fornasier2021}), although we cannot rule out other effects at present (see next paragraph).

Finally, difficulties remain in simultaneously fitting the NGA and NGT because the areas that strongly affect both in the southern hemisphere (the whole of which receives a similar amount of insolation overall) are spatiall correlated. Further splitting of activity across the surface cannot improve the fits, that is, increasing the spatial resolution of a surface activity model does not help to match the Rosetta data. This link would be broken if outgassing varied in local time over a comet rotation (i.e.~a lag angle between peak insolation and peak outgassing), suggesting that more advanced time-dependent thermal models may be necessary to fully understand the outgassing pattern of 67P and the activity mechanism of comets. In summary, both spatially and temporally varying activity is needed to fit the 67P outgassing pattern in a way that is not easily reproduced by any current thermal model.

Overall, the use of non-gravitational dynamics in the form of trajectory and rotation data clearly aids in distinguishing between different activity distributions and thermophysical models for comet 67P. This can help to test various general ideas about cometary activity and structure.

\begin{acknowledgements}
J.A. and N.A.’s contributions were made in the framework of a project funded by the European Union’s Horizon 2020 research and innovation programme under grant agreement No 757390 CAstRA. J.A. also acknowledges funding by the Volkswagen Foundation.
We thank Tobias Kramer for useful discussions and the anonymous reviewer whose comments improved the quality of this manuscript.
\end{acknowledgements}

\bibliographystyle{aa}
\bibliography{Bibliography}

\begin{appendix}

\section{Astrometry}
\FloatBarrier

\begin{table}[h]
\caption{Initial positions of 67P at $-350$ days relative to perihelion in the J2000 ecliptic coordinate frame.}
\begin{center}
\begin{tabular}{lr}
Quantity & Value \\
\hline
$t$ (Js) & 462463456.58755416 \\
$x$ (km) & $1.99549521\times10^{+08}$ \\
$y$ (km) & $-4.76677235\times10^{+08}$ \\
$z$ (km) & $-5.66149293\times10^{+07}$ \\
$\dot{x}$ (km s$^{-1}$) & $7.34031872\times10^{+00}$ \\
$\dot{y}$ (km s$^{-1}$) & $1.41777157\times10^{+01}$ \\
$\dot{z}$ (km s$^{-1}$) & $4.26145500\times10^{-01}$ \\
\end{tabular}
\label{table2}
\end{center}
\end{table}

\FloatBarrier
\section{Model F, detailed results}

\begin{figure}[h]
\resizebox{\hsize}{!}{\includegraphics{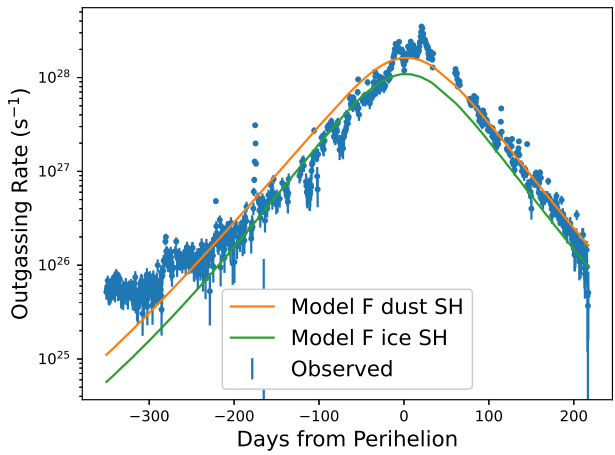}}
\caption{Observed total gas production (Rosetta/ROSINA values from \citealp{Hansen}) compared to two versions of model F, based on \citet{Fulle2020}.}
\label{Plot_resultsFQ}
\end{figure}

\begin{figure}[h]
\resizebox{\hsize}{!}{\includegraphics{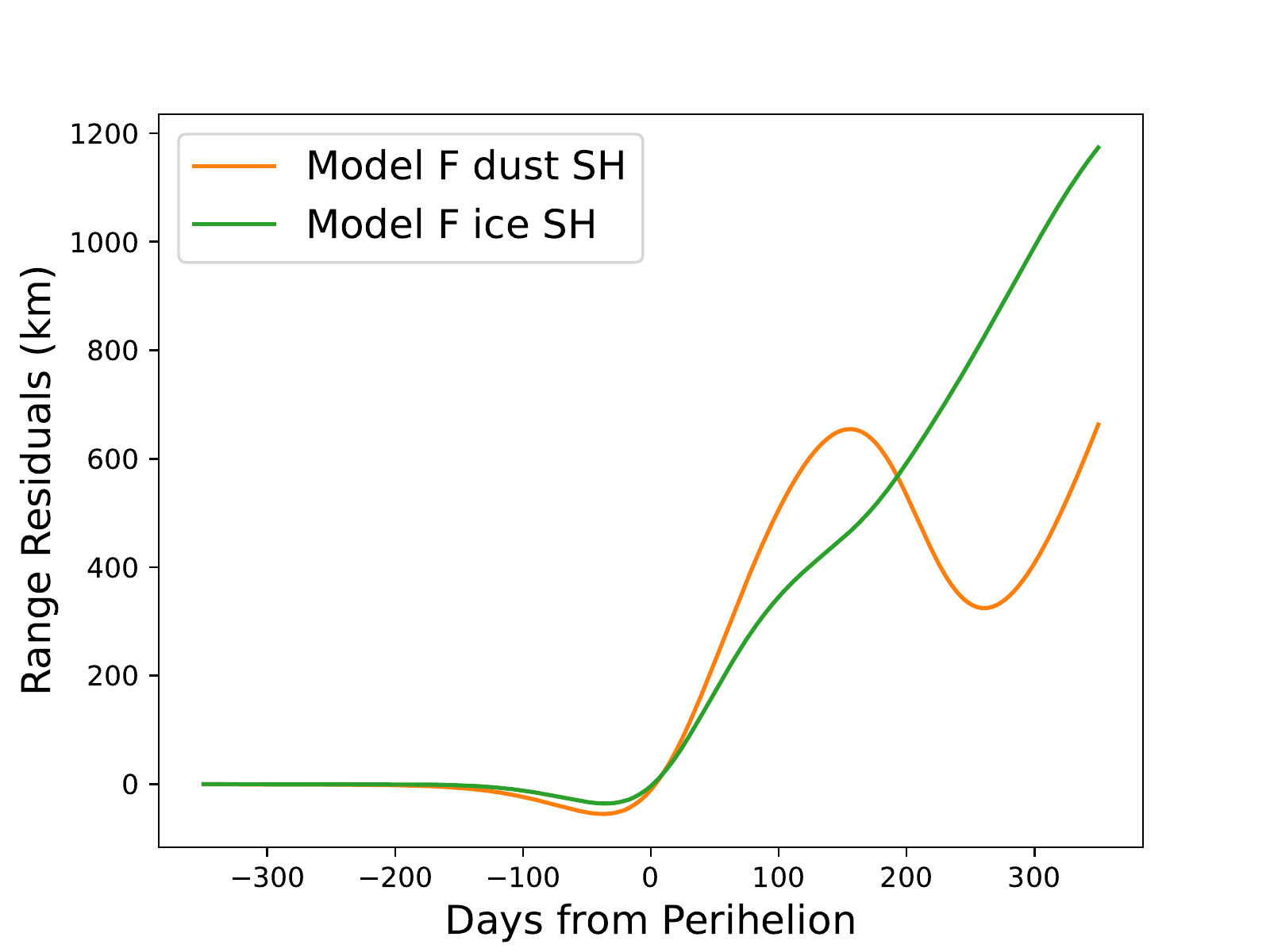}}
\caption{Observed minus computed Earth-comet range, $R$, for two versions of model F.}
\label{Plot_resultsRF}
\end{figure}

\begin{figure}[h]
\resizebox{\hsize}{!}{\includegraphics{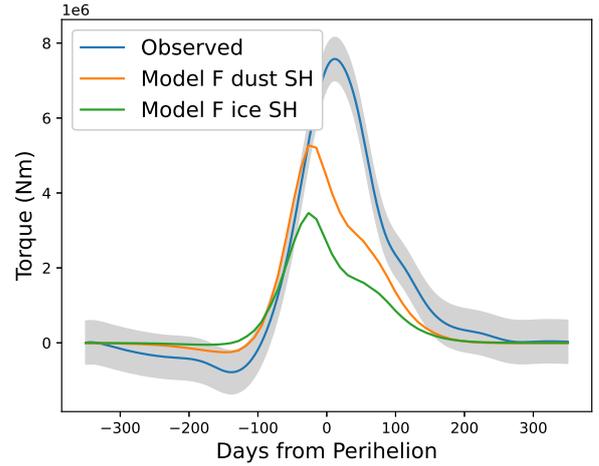}}
\caption{Observed $z$ component of the torque compared to two versions of model F.}
\label{Plot_resultsTorqueF}
\end{figure}

\begin{figure}[h]
\resizebox{\hsize}{!}{\includegraphics{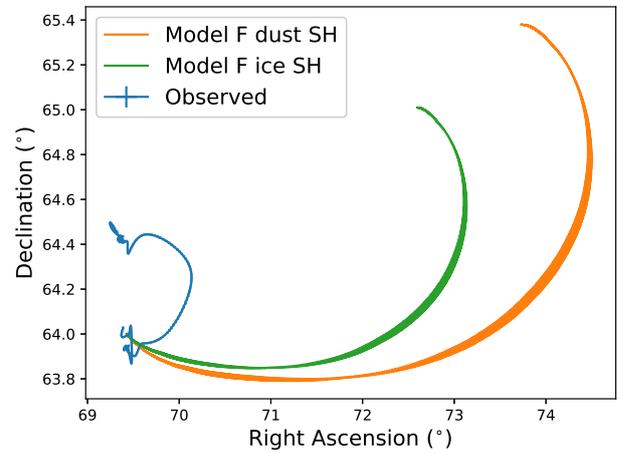}}
\caption{Observed pole orientation (Ra/dec) compared to two versions of model F.}
\label{Plot_resultsRAdecF}
\end{figure}

\begin{figure}[h]
\resizebox{\hsize}{!}{\includegraphics{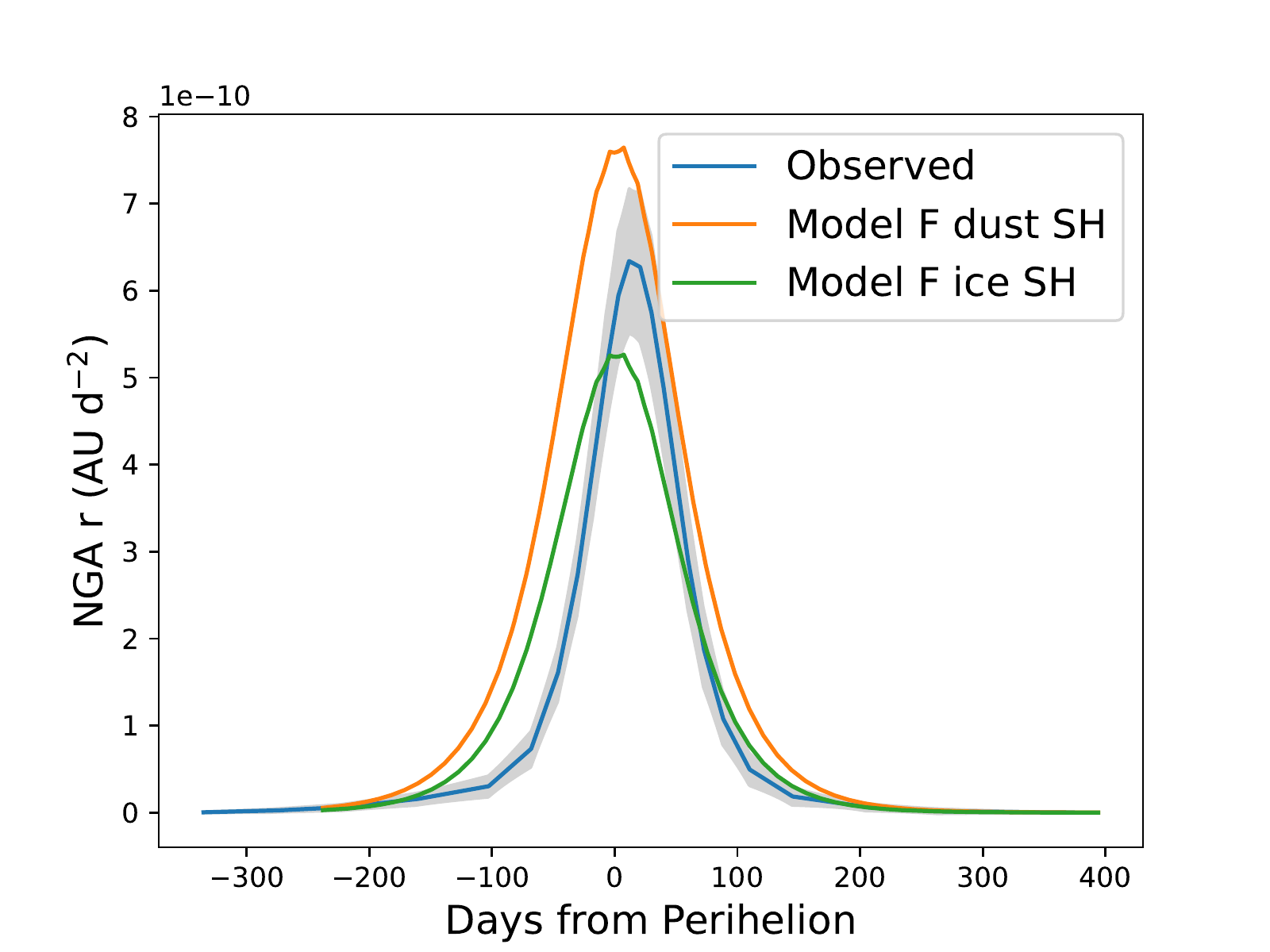}}
\caption{Observed radial acceleration in the cometary \textbf{$(\hat{r},\hat{t},\hat{n})$} frame compared to two versions of model F.}
\label{Plot_NGArF}
\end{figure}

\begin{figure}[h]
\resizebox{\hsize}{!}{\includegraphics{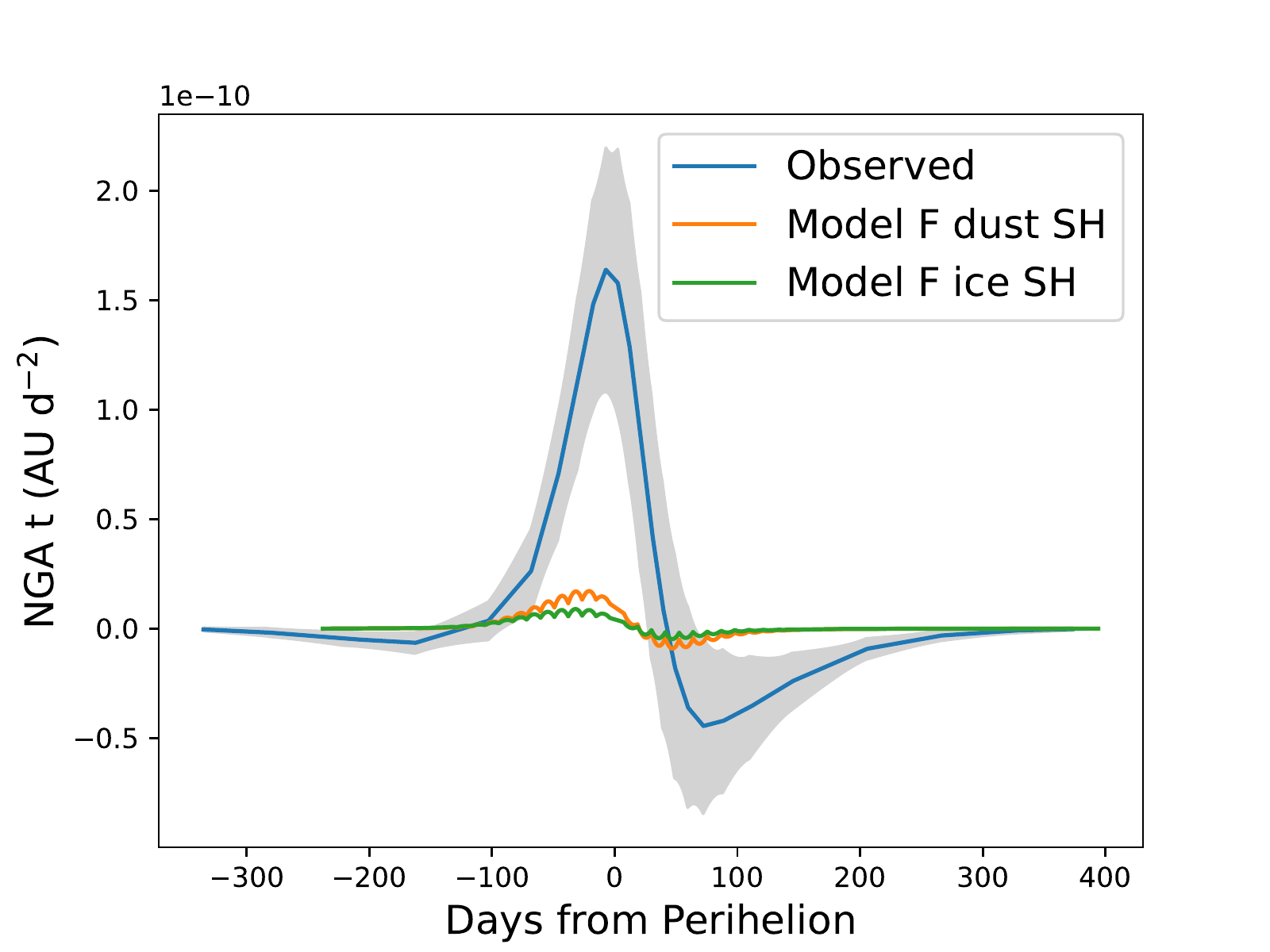}}
\caption{Observed tangential acceleration in the cometary \textbf{$(\hat{r},\hat{t},\hat{n})$} frame compared to two versions of model F.}
\label{Plot_NGAtF}
\end{figure}

\begin{figure}[h]
\resizebox{\hsize}{!}{\includegraphics{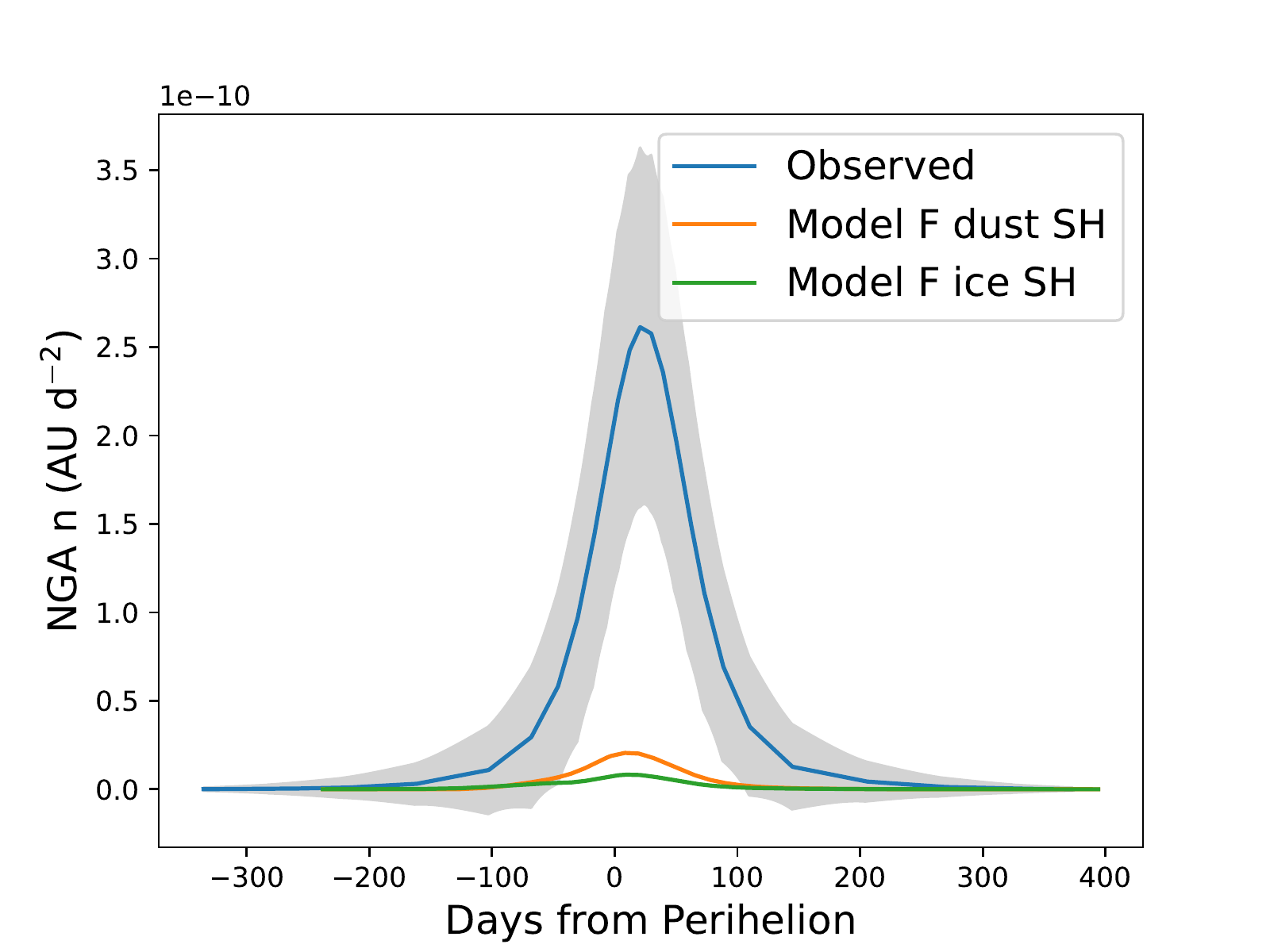}}
\caption{Observed normal acceleration in the cometary \textbf{$(\hat{r},\hat{t},\hat{n})$} frame compared to two versions of model F.}
\label{Plot_NGAnF}
\end{figure}

\end{appendix}

\end{document}